\documentclass[%
 reprint,
 amsmath,amssymb,
 aps,
]{revtex4-1}

\usepackage{graphicx,subfigure}
\usepackage{dcolumn}
\usepackage{bm}

\begin{document}

\preprint{APS/123-QED}

\title{Solid-solid collapse transition in a two dimensional model molecular system}

\author{Rakesh S. Singh}
\author{Biman Bagchi}
\email{Corresponding author: bbagchi@sscu.iisc.ernet.in}
\affiliation{ Solid State and Structural Chemistry Unit, Indian Institute of Science, Bangalore 560012, India}%

\date{\today}

\begin{abstract}
Solid-solid collapse transition in open framework structures is ubiquitous in nature. The real difficulty in 
understanding detailed microscopic aspects of such transitions in molecular systems arises from the interplay 
between different energy and length scales involved in molecular systems, often mediated through a solvent. 
In this work we employ Monte Carlo (MC) simulations to study the collapse transition in a model molecular system interacting 
via both isotropic as well as anisotropic interactions having different length and energy scales. The model we use 
is known as Mercedes-Benz (MB) which for a specific set of parameters sustains three solid phases: honeycomb, oblique 
and triangular. In order to study the temperature induced collapse transition, we start with a metastable honeycomb 
solid and induce transition by heating. High density oblique solid so formed has two characteristic length scales 
corresponding to isotropic and anisotropic parts of interaction potential. Contrary to the common believe and 
classical nucleation theory, interestingly, we find linear strip-like nucleating clusters having significantly 
different order and average coordination number than the bulk stable phase. In the early stage of growth, 
the cluster grows as linear strip followed by branched and ring-like strips. The geometry of growing cluster is 
a consequence of the delicate balance between two types of interactions which enables the dominance of stabilizing 
energy over the destabilizing surface energy. The nuclei of stable oblique phase are wetted by intermediate order 
particles which minimizes the surface free energy. We observe different pathways for pressure and temperature induced 
transitions.

\end{abstract}

\maketitle


\section{Introduction}
Several studies have shown that many solids exhibit temperature and pressure induced solid-solid phase 
transitions resulting in a discontinuous change in material properties\cite{1,2,3,4,5,6,7}. In the case of shock wave 
induced solid-solid transitions both temperature and pressure of the system increases\cite{4,5}. The well known 
examples of solid-solid transition are Fe and its alloys (Fe/Ni alloys transform on cooling from face-centered cubic 
(fcc) phase to a low-temperature body-centered cubic (bcc) phase), alloys based on CuAl, NiTi, NiAl, 
and oxide ceramics (like ZrO2), name only a few. In steel, the structural transitions are used to enhance 
its strength. Carbon crystallizes in the hexagonal graphite structure at ambient pressure and undergoes 
a pressure-induced transition to the tetrahedral diamond structure. Structural transformations are also 
observed in biological systems. Some virus species use the pressure-induced martensitic transformation to 
infect bacteria cells\cite{1}. Another application of solid-solid transition is the shape-memory effect in many 
alloys and widely used in medicinal science\cite{3}. Solid-solid transitions can be driven either by homogenous or 
heterogeneous (in the case of presence of preexisting defects which reduces the free energy barrier for transition) 
nucleation. Despite the great importance of solid-solid transitions in material, medicinal and biological sciences, 
little is known about the microscopic aspects of solid-solid transitions. In molecular systems, the real difficulty 
in understanding detailed microscopic aspects of such transitions arises from the interplay between different 
energy and length scales involved and often mediated through a solvent. \\
Computer simulations are the most appropriate technique to study the microscopic aspects of solid-solid transitions. 
However, the difficulty in computer simulations arises in construction of potential which can give 
rise to multiple solid phases. Now it is well accepted that systems interacting with core soft 
potentials show rich phase diagram consisting multiple solid phases\cite{8,9,10,11,12}. Core-softened potential 
was first used by Hemmer and Stell\cite{8} in a lattice gas system to discuss the isostructural solid-solid phase 
transition. Later, Young and Alder\cite{9} showed that a two dimensional system interacting via repulsive step 
potential (hard-core plus square-shoulder repulsion) can rise to phase diagram and solid-solid phase 
transitions qualitatively similar to those observed in cesium and cerium. Distinct solid phases in systems 
interacting with core soft isotropic potential arise due to the presence of distinct length scales in the 
potential. Recently, Sengupta and coworker\cite{13} introduced a potential where particles interact via modified 
Lennard-Jones potential and each particle has two internal states. They also observed a rich phase diagram 
showing water-like thermodynamic anomalies and polymorphism. Although there are many computer simulation studies 
that discuss the rich phase diagram consisting many solid phases\cite{8,9,10,11,12,13,14} only few studies are devoted for 
understanding of nature and pathway of solid-solid transitions. Recently, Rao and Sengupta\cite{15,16,17} studied 
the nucleation of a triangular phase from bulk metastable square crystal in a model two dimensional system 
interacting via both two and three body potentials. These studies suggest that at high temperature the critical 
nucleus of the triangular phase is isotropic and at low temperature the critical nucleus is anisotropic twinned 
bicrystal. They also observed that the final microstructures depend on quench rate. For systems like iron the 
dependence of microstructures on rate of cooling (quench depth) is well known. Slow cooling rate results in an 
equilibrium ferrite phase while rapid cooling results the trapping of system in a metastable twinned martensite. \\
In this work we have studied the microscopic pathway for the solid-solid collapse transition in a model two dimensional 
(2D) molecular system interacting via both isotropic as well as anisotropic interactions having different length and 
energy scales. The model we use is known as Mercedes-Benz (MB) potential\cite{18,19,20} which on specific parameterization 
for relative strength of isotropic interaction sustains three solid phases: honeycomb, oblique and triangular\cite{21}. 
The complexity of MB model and simplicity of 2D allows us to explore several aspects of solid-solid transitions not 
observed in atomic systems. In the subsequent sections we shall discuss the honeycomb to oblique solid 
collapse transition. \\
The organization of the rest of the paper is as follows. In the next section (Section II), we present the model 
system and the simulation details used in this work. In Sections III, first, we discuss the 
free energy surface and then the microscopic pathway for temperature and pressure induced honeycomb 
to oblique solid transition. In Section IV we present concluding remarks.

\section{Model and Simulation Details}
Mercedes-Benz potential has two terms - isotropic Lennard-Jones (LJ) and an anisotropic 
hydrogen bond (HB) term written in a completely separable form as \cite{18,19,20},
\begin{equation}
U(X_i,X_j) = U_{LJ}(r_{ij}) + U_{HB}(X_i,X_j),
\end{equation}
where $X_i$ denotes vectors representing both the coordinates and orientation of the $i^{th}$ particle and $r_{ij}$ 
the distance between the centers of two molecules. $U_{LJ}$ is defined as 
 \begin{equation}
U_{LJ}(r_{ij}) = 4\varepsilon_{LJ}[(\sigma_{LJ}/r_{ij})^{12}-({\sigma_{LJ}/r_{ij}})^6],
\end{equation}  
where $\varepsilon_{LJ}$ and $\sigma_{LJ}$ are the wel-depth and diameter for the isotropic LJ interaction. 
The anisotropic HB part,$U_{HB}$, is defined as 
\begin{equation}
\begin{split}
 U_{HB}(X_i,X_j) = \varepsilon_{HB}G(r_{ij}-\sigma_{HB}) \\
       \times\sum_{k,l=1}^3G(i_k.u_{ij}-1)G(j_l.u_{ij}+1),
\end{split}
\end{equation}
where $G(x) = exp\left(-x^{2}/2\sigma^{2}\right)$ with $\sigma$ = 0.085. 
The unit vector $i_{k}$ represents the \textit{k${}^{th}$} arm of the \textit{i${}^{th}$} particle 
(\textit{k} = 1, 2, 3) and $u_{ij}$ is the unit vector joining the center of particle \textit{i} 
to the center of particle \textit{j}. The HB parameters are $\varepsilon_{HB}=-1$ and bond 
length $\sigma_{HB}=1$. LJ contact distance ($\sigma_{LJ}$) is $0.7\sigma_{HB}$. 
Anisotropic parameter which is a measure of relative strength of the isotropic to anisotropic interaction 
is defined as $\lambda_{i}=\varepsilon_{LJ}/\varepsilon_{HB}$. In the present study 
we have used $\lambda_{i}=0.4$. Note that we have two length as well as energy scales corresponding 
to the isotropic LJ and anisotropic HB interactions. Monte Carlo simulations are performed in constant pressure 
and temperature (NPT) ensemble having number of particles $N$ = 576, 836 and 1008. All quantities are represented 
in reduced units scaled with HB parameters such as temperature as 
${k_{B}T\mathord{\left/{\vphantom {k_{B}T\varepsilon_{HB}}} \right.\kern-\nulldelimiterspace}|\varepsilon_{HB}}|$ 
and distance is scaled with hydrogen-bond length, $\sigma_{HB}$.\\
In the case of temperature induced transition (except Fig. 1C) we started with honeycomb structure at 
$T$ = 0.05 and then increased the temperature in steps of 0.01 up to $T$ = 0.10 and then increased in steps 
of 0.0005 at $P$ = 0.10. We first equilibrated the system for $5\times10^{6}$ MC steps and then 
collected the trajectories for next $5\times10^{6}$ MC steps. For pressure induced transition 
we started with honeycomb lattice at pressure $P$ = 0.10 and then increased the pressure in steps of 
0.05 up to $P$ = 0.4 and then increased in steps of 0.01 at $T$ = 0.05. 
In this case the equilibration step was $1.5\times10^{7}$ MC steps and then collected the trajectories 
for next $1\times10^{7}$ MC steps.

\section{Results and Discussion}
\subsection{Free energy surface}
As the form of the Mercedes-Benz potential is written as a completely separable isotropic and anisotropic 
parts having distinct ground states (triangular and honeycomb), one would expect that the interplay between 
the relative strength of the isotropic to anisotropic interaction can give rise distinct solid phases 
-- either stable or metastable. Recently, we find that the interplay between the strength of 
isotropic and anisotropic interactions gives rise to rich phase diagram\cite{21}. The computed phase diagram 
consists of isotropic liquid and three crystalline phases -- honeycomb, oblique and triangular. 
Interestingly, we note that the oblique phase is neither the ground state of isotropic nor the anisotropic 
part of the potential. It forms due to delicate balance between the two types of interactions. In order to study 
the solid-solid collapse transition from low density open structured honeycomb to high density oblique phase 
the honeycomb solid (ground state of the anisotropic part of the potential) should be metastable 
(as shown in Fig. 1A) at a given temperature, pressure and relative strength of isotropic 
interaction ($\lambda_{i}$). We have constructed such a free energy surface by tuning the relative 
strength of the isotropic interaction ($\lambda_{i}$). \\
\begin{figure*}
\includegraphics[width=0.8\textwidth]{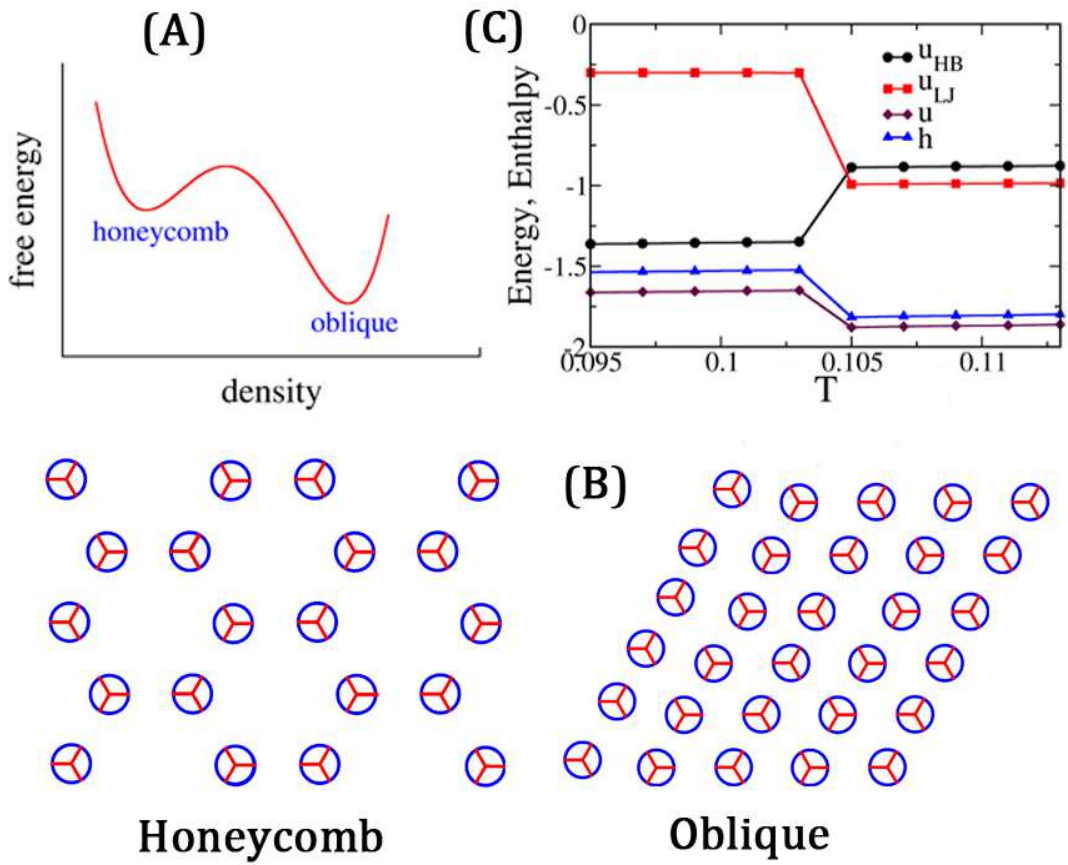}
\caption{(A) The proposed free energy surface for honeycomb to oblique solid-solid 
transition at $\lambda_{i}$ = 0.4 is shown. (B) The structures of the 
honeycomb and oblique solid phases are shown. Alignment of bonds indicates hydrogen bonding. 
(C) Isotropic ($u_{LJ}$) and anisotropic ($u_{HB}$) interaction energies per particle and 
the total energy per particle (u) before and after transition are shown. Blue line (with triangle up symbol) 
indicates the dependence of enthalpy change per particle (h) on temperature. Note the very weak 
temperature dependence of energy/enthalpy and switch in the contributions of isotropic and anisotropic 
energies to the total energy on transition.}
\label{fig1a}
\end{figure*}
The honeycomb lattice is stabilized by 3 hydrogen bonds (HBs) per particle and the oblique phase is 
stabilized by two HBs per particle along with LJ interactions (see Fig. 1B). 
Alignment of bonds in Fig. 1B indicates hydrogen bonding. Since both the phases are solid 
and stabilized by anisotropic interactions (either completely or partially), one would expect that 
entropy of two phases will not differ significantly. Only vibrational entropy (and some contribution 
from mobility of defects) is present and also not expected to differ significantly as both the solid phases 
are stabilized by anisotropic interactions, though, the fraction of directional bonding decreases on going from 
honeycomb to oblique phase. This indicates that entropy has very little role (compared to gas-liquid and 
liquid-solid transitions) in the construction of free energy surface for solid-solid transitions. Thus the 
temperature change should not alter significantly the free energy surface; however, it is important to 
overcome the required free energy barrier for the transition. \\
To support the proposed free energy surface (Fig. 1A) in Fig. 1C, 
we have shown the temperature dependence of enthalpy, energy and relative contribution of the isotropic 
(LJ) and anisotropic (HB) interactions to the total energy per particle of the system for isotropic to 
anisotropic strength ratio, $\lambda_{i}$ = 0.4. The starting configuration is 
honeycomb lattice at pressure $P$ = 0.10 and $T$ = 0.05, which on heating collapses to the high density oblique phases. 
Although, in Fig. 1C we have shown the dependence of energy and enthalpy on temperature near 
the transition temperature only, we find that very weak dependence persists up to much lower temperature $T$ = 0.05. 
Note that, contrary to the ordinary solid-liquid transition (where total energy of the system increases), 
in this case the total energy of the system decreases. A very weak dependence of enthalpy on temperature 
and decrease in the enthalpy of the system on transition supports the free energy surface proposed in 
Fig. 1A. Honeycomb solid is metastable at given initial pressure and temperature ($T$ = 0.05 and $P$ = 0.10).
 We also note that the relative contributions of LJ and HB parts of interaction potential to the total energy per 
particle is switching. This indicates the increased importance of isotropic interaction in the oblique solid phase. \\
\begin{figure}[ht!]
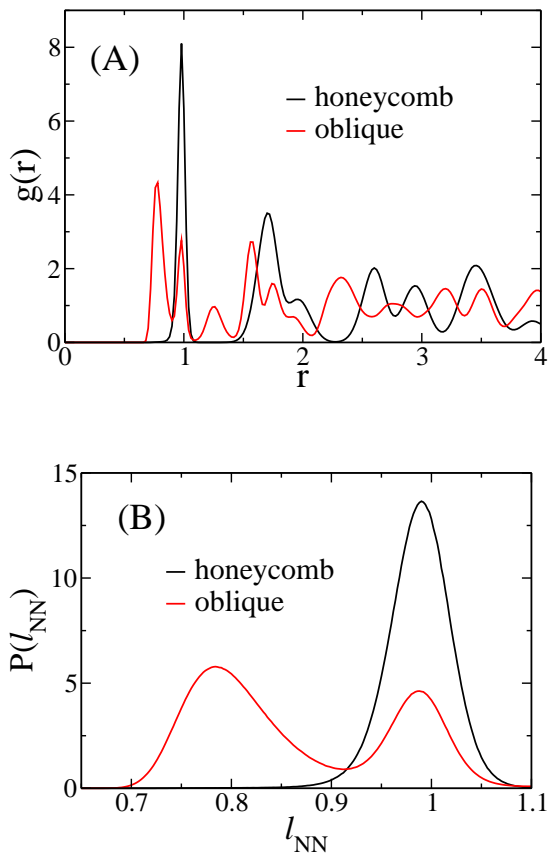

     \begin{center}
        \subfigure{
            \label{fig2a}
            \includegraphics[width=0.4\textwidth]{FIG2A.eps}
        }\\
\vspace{0.6cm}
        \subfigure{
           \label{fig2b}
           \includegraphics[width=0.4\textwidth]{FIG2B.eps}
        }\\ 
    \end{center}
    \caption{(A) Radial distribution functions for honeycomb and oblique solid phases are shown. 
(B) Bond length distributions of nearest neighbors for honeycomb and oblique solid phases are shown. 
Note the splitting of first peak of radial distribution function for the oblique and also the bimodal 
bond length distribution of nearest neighbors. The splitting of the 1${}^{st}$ peak of radial 
distribution function and bimodal distribution of nearest neighbor bond length distribution suggest 
that the oblique solid has two characteristic length scales corresponding to hydrogen-bond and vander 
Waals (LJ) interactions.}
   \label{fig2}
\end{figure}
In order to gain quantitative insight into the structures of honeycomb and oblique phases, in Fig. 2 
we have shown the computed radial distribution functions and bond length distributions of nearest neighbors 
for both honeycomb and oblique solid phases near transition temperature. We have used the Voronoi 
construction in order to identify the nearest neighbors ($NN$) of each particle. Note the splitting of 
first peak of radial distribution function and also the bimodal bond length distribution of nearest neighbors 
for the oblique solid phase. This signifies that in the oblique solid all nearest neighbor particles are not equidistant. 
There are two length scales present in the system, which is a consequence of the two length scale potential. 
Particles having lower bond length are stabilized by isotropic LJ interaction and larger bond lengths are 
stabilized by anisotropic HB interaction. Contrary to the oblique solid, honeycomb solid is stabilized only 
by anisotropic HB interactions and thus the nearest neighbor distance distribution has only one peak centered 
at HB length and there is no splitting in the first peak of radial distribution function. \\
In the subsequent sections, first, we shall discuss the temperature induced transition followed by a comparative study 
of pressure and temperature induced transitions.
\subsection{Order parameters and their susceptibilities} 
In Fig. 3, we have shown the order parameters change and their susceptibilities for temperature induced 
collapse transition. On increasing temperature, thermal fluctuations destabilize the honeycomb 
structure and collapses to a high density oblique structure with large change in density 
(from $\sim{0.79}$ to $\sim{1.56}$) (see Fig. 3A). In the collapsed transition, high density change indicates 
that the transition can only be first order. As due to increase in order parameter difference the surface 
energy cost for formation of new phase also increases. Since the barrier in the global free energy surface 
as function of global order parameter (density and order) is related to the surface tension, the probability 
of long wavelength fluctuations of the global order parameters (density fluctuation is coupled with order fluctuation) 
in the system is less. In Fig. 3B, we have reported the change of both the global six-fold bond 
orientational order ($\psi_{6g}$) and local three-fold bond orientational order ($\psi_{3l}$) with 
temperature. \textit{m}-fold global bond orientational order ($\psi_{mg}$) is defined as
\begin{equation} \label{4)} 
\psi_{mg}=\frac{1}{N}\left\langle\left|\sum_{k=1}^{N}\left(\,\frac{1}{NN} 
\sum_{j=1}^{NN}e^{im\theta_{kj}}\right)\,\right|\right\rangle,         
\end{equation} 
and \textit{m}-fold local bond orientational order ($\psi_{ml}$) is defined as 
\begin{equation} \label{5)} 
\psi_{ml}=\frac{1}{N}\left\langle\sum_{k=1}^{N}\left(\,\left|\frac{1}{NN} 
\sum_{j=1}^{NN}e^{im\theta_{kj}}\right|\right)\right\rangle,        
\end{equation} 
where \textit{m} = 6 for six-fold and \textit{m} = 3 for three-fold symmetry. \textit{NN} is the 
number of nearest neighbors, $\theta_{ij}$ is the angle of the bond that the \textit{j${}^{th}$} 
neighbor makes with the tagged \textit{i}${}^{th}$ particle. Significantly large values of 
global order parameter,$\psi_{6g}$, in both phases indicate that there is a global translation 
of either six-fold or three-fold bond orientational order. As $\psi_{6g}$ does not distinguish between 
the six-fold and three-fold bond orientation order, we have also plotted the local three-fold bond 
orientational order (note that global three-fold bond orientational order is zero for both six-fold 
and three-fold symmetry). However, the later does not guarantee the long range translation of local 
bond orientational order. Sharp transition in $\psi_{3l}$ and significantly large values of $\psi_{6g}$ 
in both the phases indicate a one step transition from three-fold honeycomb lattice to six-fold (distorted) 
oblique solid phase.\\
\begin{figure*}[ht]
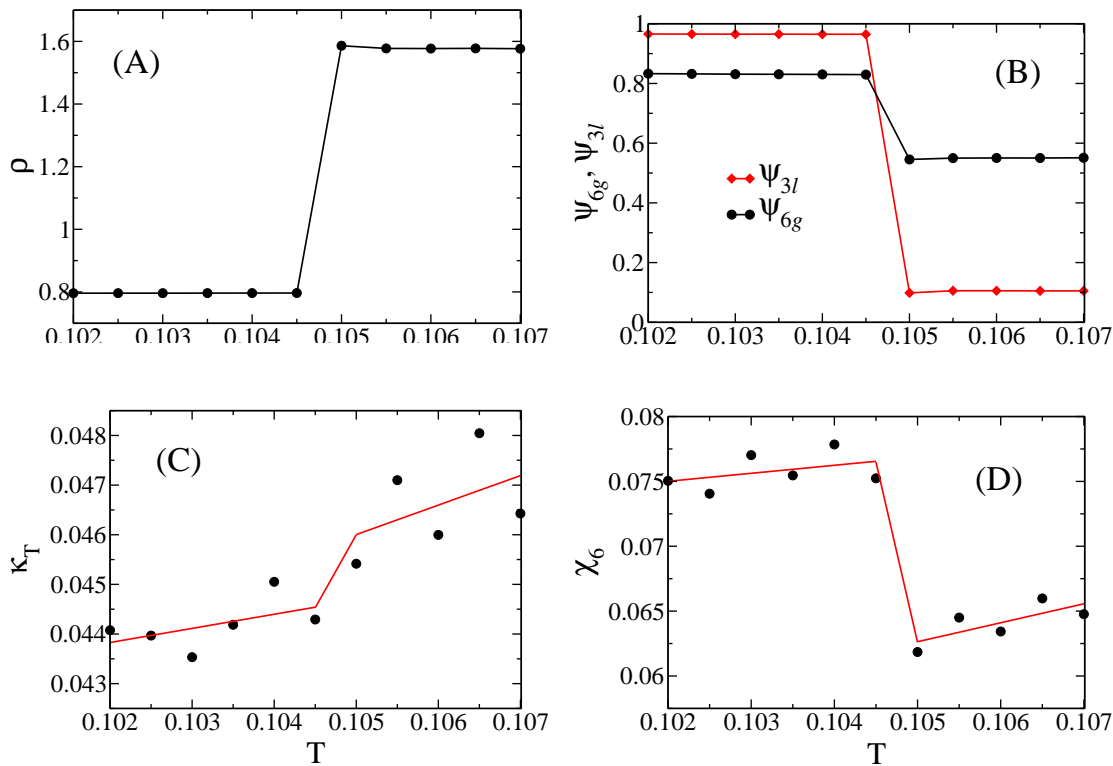

     \begin{center}
        \subfigure{
            \label{fig3a}
            \includegraphics[width=0.4\textwidth]{FIG3A.eps}
        }
        \subfigure{
           \label{fig3b}
           \includegraphics[width=0.4\textwidth]{FIG3B.eps}
        }\\ 
        \subfigure{
            \label{fig3c}
            \includegraphics[width=0.4\textwidth]{FIG3C.eps}
        }
        \subfigure{
            \label{fig3d}
            \includegraphics[width=0.4\textwidth]{FIG3D.eps}
        }
    \end{center}
    \caption{(A) The dependence of the density ($\rho$) of the system on temperature is shown. 
The starting configuration is honeycomb solid and after transition it goes to the oblique phase. 
(B) The dependence of global six-fold ($\psi_{6g}$) and local three fold ($\psi_{3l}$) bond orientational 
order parameters on temperature is shown. In (C) and (D), we have shown the dependence susceptibilities 
of global order parameters (density and six-fold bond orientational order parameter): isothermal 
compressibility ($\kappa_{T}$) and six-fold bond orientational order susceptibility ($\chi_{6}$) on 
temperature. Weak temperature dependence and finite jump in response functions suggests strongly 1${}^{st}$ 
order transition.}%
   \label{fig3}
\end{figure*}
In Fig. 3C and 3D, we have plotted the susceptibilities along density 
(isothermal compressibility, $\kappa_{T}$) and global six-fold bond orientational order parameter 
$\left(\chi_{6}\right)$. Six-fold bond orientational order susceptibility is defined 
as $\chi_6 = \langle|{\psi_{6g}}|^2\rangle - \langle|{\psi_{6g}}|\rangle^2$ and and isothermal compressibility is 
defined in terms of volume fluctuations as $\kappa_T=({\langle{V^2}\rangle-\langle{V}\rangle^2})/{T\langle{V}\rangle}$. 
Weak temperature dependence and finite jump in both $\kappa_{T}$ and $\chi_{6}$ 
suggest the strongly 1${}^{st}$ order transition.

\subsection{Microscopic pathway of temperature induced collapse transition}
In order to find the pathway of transition first we need to distinguish the particles belonging to 
new phase in the sea of metastable bulk honeycomb solid phase. To distinguish the nucleating and 
growing new oblique-like particles from the parent metastable honeycomb phase, we computed the distribution 
of local 3-fold bond orientational order ($\psi_{3i}$),  
defined as $\psi_{3i}=\left|{\sum_{j=1}^{NN}e^{i3\theta_{ij}}\mathord{\left/ 
{\vphantom{\sum_{j=1}^{NN}e^{i3\theta_{ij}}NN}}\right.\kern-\nulldelimiterspace}NN}\right|$, 
for both the solid phases. As shown in Fig. 4, the complete separation of two distributions 
\begin{figure}
\includegraphics[width=0.45\textwidth]{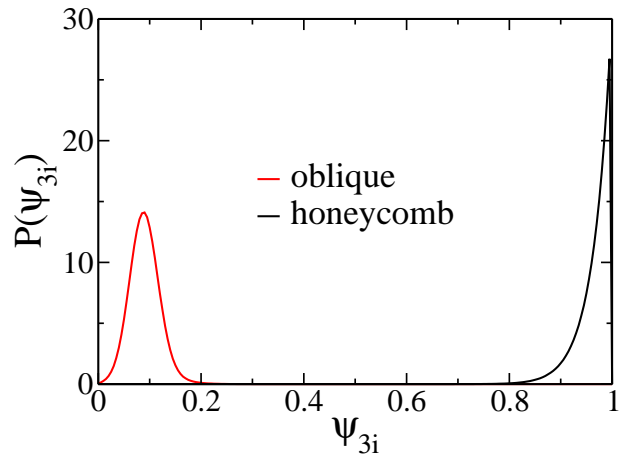}
\caption{Distributions of 3-fold local bond orientational order ($\psi_{3i}$) for honeycomb and 
oblique solids are shown.}
\end{figure}
\begin{figure*}
\includegraphics[width=0.8\textwidth]{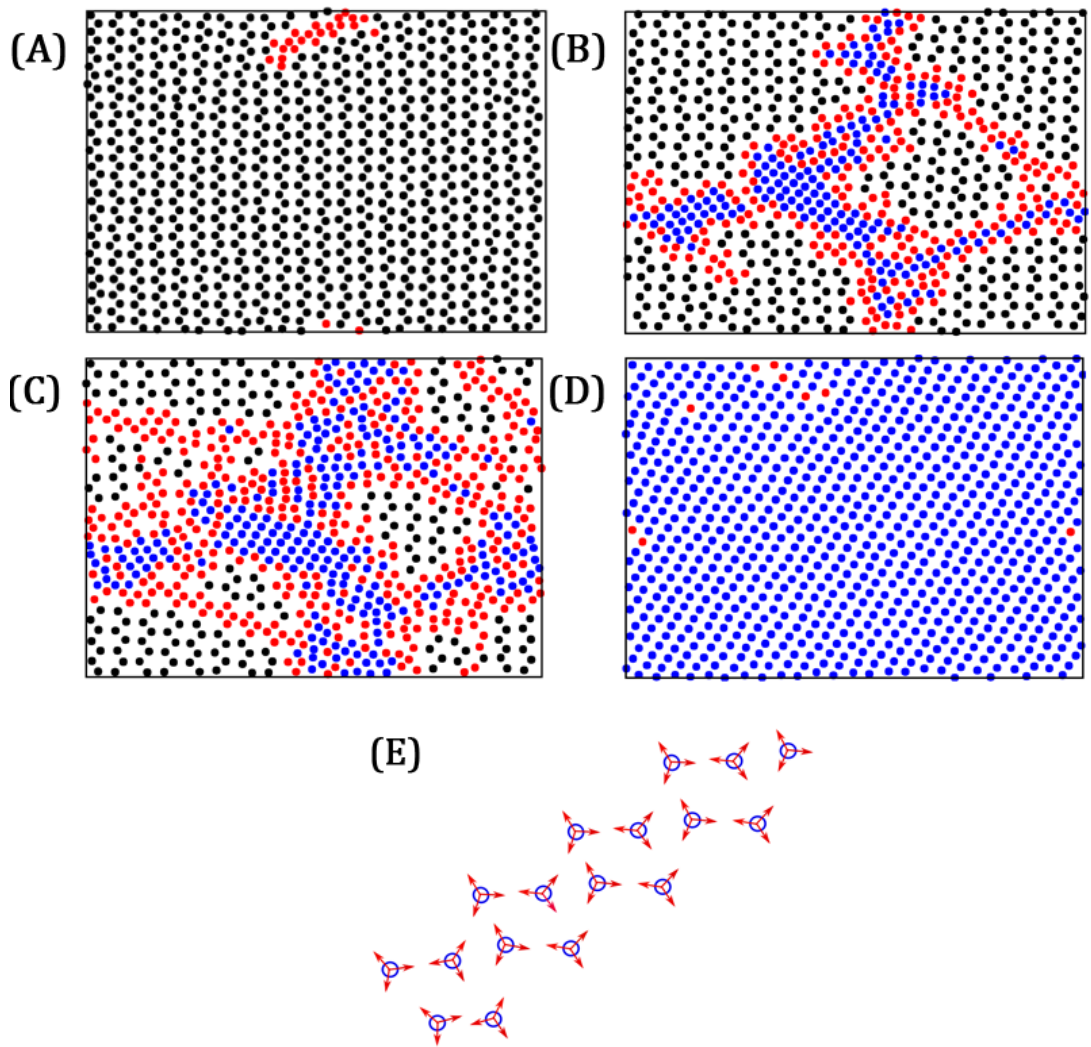}
\caption{Snapshots of the system at (A) early (B) intermediate and (C) late stage of 
transition are shown. (D) Oblique solid phase after transition for $N$ = 836 particles are shown. 
Note the linear strip-like arrangement transforming into branched and ring-like strips. Oblique solid 
has string-like arrangement of particles. We also note the wetting of oblique phase by a layer of 
intermediate solid phase. (E) Snapshot of a growing nucleus is shown.}
\end{figure*}
enables us to separate the growing new phase from the parent honeycomb solid phase. Due to significant overlap 
in the distribution of local six-fold bond orientational order we have not used it as a parameter to distinguish 
the two phases. Note that, though, $\psi_{3i}$ is able to separate the oblique solid from the 
honeycomb solid phase, it does not distinguish the oblique solid-like and liquid-like particles 
(as the peak of distribution is centered at very low value of $\psi_{3i}$). 
A two order parameter criteria using the values of both $\psi_{3i}$ and  $\psi_{6i}$ is required 
to separate the oblique phase with liquid-like particles as the oblique solid is characterized with 
low $\psi_{3i}$ and high $\psi_{6i}$ values (see Fig. 3B).\\
In Fig. 5A - 5D, we have shown the snapshots of the system at different stages during the 
collapse transition. Depending on $\psi_{3i}$ value we have defined three types of 
particles in the system. The particles having $\psi_{3i}$ value greater than 0.75 is 
considered as honeycomb type and indicated with black color, red color particles have intermediate order 
with $\psi_{3i}$ value between 0.25 and 0.75. The particles having $\psi_{3i}$ 
value less than 0.25 and $\psi_{6i}$ value greater than 0.2 (lower cut-off value of $\psi_{6i}$ below 
which the distribution of $\psi_{6i}$ $\left(P(\psi_{6i})\right)$ in oblique phase is zero) are 
considered as oblique-like. Note that these criteria do not distinguish oblique-like and triangular solid-like 
particles. However, simplicity of 2D systems allows us to distinguish these two types of orders by 
visual inspection of snapshots. Recently, Onuki \textit{et} \textit{al}. developed an elegant approach based 
on local bond orientational order parameter to quantify different phases, stacking faults and grain boundaries 
in both two and three dimensional systems\cite{22,23}.\\
\begin{figure*}
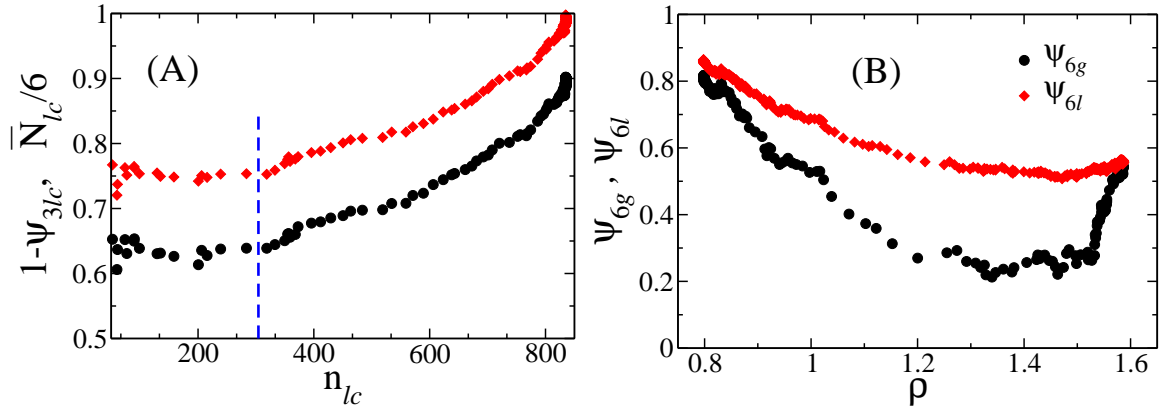

     \begin{center}
        \subfigure{%
            \label{fig6a}
            \includegraphics[width=0.42\textwidth]{FIG6A.eps}
        }%
        \subfigure{%
           \label{fig6b}
           \includegraphics[width=0.42\textwidth]{FIG6B.eps}
        }\\ 
    \end{center}
    \caption{(A) The dependence of local order parameter, 1-$\psi_{3lc}$ (black filled circles) 
and average coordination number scaled with six (the average coordination number of 
particles in oblique solid phase), $\bar{N}_{lc}/6$ (red filled diamonds) of the largest 
cluster on the size of the largest cluster is shown. For convenience in comparison 
we have plotted $1-\psi_{3lc}$ in place of $\psi_{3lc}$. All quantities are averaged over 
100 MC steps along the growth trajectory for $N$ = 836 particle system. We have also observed 
similar trend for $N$ = 1008. The blue dotted line indicates the critical cluster size up to 
which there is no appreciable change in either average order or the coordination number of the cluster. 
(B) Pathways of solid-solid transition in two order parameter (density and six-fold bond orientational order) 
space are shown. The black filled circles represent the evolution of global six-fold bond orientational 
($\psi_{6g}$) order with density and red filled diamonds represent the evolution of local six-fold 
bond orientational order ($\psi_{6l}$) with density.}
   \label{fig6}
\end{figure*}
 Contrary to the common believe and classical nucleation theories that the nuclei of new phase are circular 
in 2D) and have same order and density as of bulk stable phase, interestingly, we find highly anisotropic growth 
of new phase. At the early stage of growth, the cluster grows as linear strip (having significantly 
different order than the bulk oblique phase) followed by branched and ring-like strip and finally 
engulfs the whole system. The geometry of nucleating and growing cluster (shown in Fig. 5E) 
is a consequence of delicate balance between the two (isotropic and anisotropic) types of interactions 
which enables the dominance of stabilizing energy over the destabilizing surface energy. Sharing of 
the H-bonds (see Fig. 5E) between the growing cluster and the bulk metastable honeycomb phase 
minimizes the surface energy cost for formation of new phase. We also observe that the structure of the growing 
cluster is different from the bulk oblique solid phase. As shown in Fig. 5B and 5C, 
the oblique-like phase starts appearing inside the high density intermediate order cluster. There is no 
clear dividing surface between the oblique and the honeycomb and hence it is difficult to assign unambiguously 
the identity of particles as honeycomb or oblique-like. Wetting of the oblique-like phase by intermediate 
order particles reduces the surface free energy cost for the formation of stable oblique phase.\\
Thus, the solid-solid transition in a molecular system is a two step process. In the first step small 
anisotropic cluster having significantly different order nucleates and grows. 
Anisotropic shape enables the cluster to maximize the energy gain due to strong specific interactions 
among particles inside the cluster as well as with the bulk parent honeycomb phase. In the second step, 
due to internal rearrangement inside the dense (intermediate ordered) anisotropic cluster oblique-like 
phases start appearing. Wetting of nuclei by intermediate order phase is also observed in crystal 
nucleation from metastable liquid\cite{24} where nucleation preferentially takes place in regions of 
high structural order. Note that clear dividing surface with no sharing of particles between the two 
phases (cluster of stable phase and bulk metastable phase) creates a density/order deficiency and thus increases 
the surface energy\cite{25}. Wetting is one of the factors responsible for the observed discrepancy 
between the predictions of classical nucleation theory and experimental/simulation studies.\\
In Fig. 6A, we have shown the evolution of the average order and average coordination number of 
a growing cluster. We consider a particle not as honeycomb like if it has $\psi_{3i}$ value less than 0.75. 
Particles having $\psi_{3i}$ value less than 0.75 and are connected by neighborhood form a cluster of 
new phase. The order of the cluster ({$\psi _{3lc}$) is defined as, 
\textbf{$\psi_{3lc}={\sum_{i=1}^{n_{lc}}\psi_{3i}\mathord{\left/ 
{\vphantom{\sum_{i=1}^{n_{lc}}\psi_{3i}n_{lc}}}\right.\kern-\nulldelimiterspace} n_{lc} }$,} 
where $n_{lc}$ is the size of the largest cluster and summation is over all the particles that 
belong to the largest cluster. Average coordination number is scaled with the average coordination number 
of particles in the oblique solid phase. Since only one cluster grows in the system the largest cluster 
contains the maximum fraction of particles belonging to the new phase. The cluster of new phase has 
significantly different order from the bulk stable phase. During the growth, both the order 
and average coordination number of the cluster gradually evolve and finally merge to the value 
corresponding the bulk oblique solid phase. As shown in the figure, there is a crossover 
(indicated by blue dotted line in Fig. 6A) from weakly dependence to non-linear dependence of order
 as well as average coordination number on size of the cluster. At early stage of growth, weak dependence 
of the order and average coordination number on size of the cluster indicates the linear strip-like growth, 
where surface to volume ratio does not change appreciably. The non-linear dependence of order/average coordination 
number on cluster size is a consequence of two factors: (i) the change of surface to bulk ratio due to branching 
and swelling of the growing cluster and (ii) the rearrangement of the particles inside the cluster towards 
oblique-like structure. \\
In Fig. 6B, we have shown the evolution of both global six-fold bond orientational order 
($\psi_{6g}$) and local six-fold bond orientational order 
($\psi_{6l}$) of system with density during collapse transition. 
We observe an almost monotonic decrease of $\psi_{6l}$ during transition from 
the value in the honeycomb to its value in the oblique solid phase. However, $\psi_{6g}$ 
shows strongly non-monotonic evolution from its value in the honeycomb to the oblique solid phase. 
This non-monotonic evolution of $\psi_{6g}$ and monotonic (or very weakly non-monotonic) 
evolution of $\psi_{6l}$ arises due to cancellation of components of global bond 
orientational order at intermediate stage of growth and suggests the presence of domains 
(separated by grain boundaries) having different symmetry axes. Presence of domains having 
distinct symmetry axes does not alter the values of local order parameters. This scenario is also 
evident from the snapshots shown in Fig. 5. In the late stage of growth due to relaxation 
of grain boundaries the global order parameter value again increases and finally merges to the value 
corresponding to the bulk oblique phase.
\subsection{Does the pathway of pressure induced transition differ from temperature induced transition?}
Temperature plays two roles in the phase transition -- (i) it couples with entropy and perturbs the 
entire free energy landscape and (ii) it provides sufficient thermal energy to overcome the free energy 
barrier for phase transition. In the case of condensation and crystallization, where the entropy 
difference between the two phases is large, the dependence of free energy barrier on temperature 
is stronger than the case of solid-solid phase transitions. As discussed earlier, in the case of 
solid-solid transition where specific interactions play important role in the stability of both 
the phases we can assume that the major role of temperature is to provide sufficient thermal 
energy to overcome the free energy barrier for transition. As the honeycomb solid is metastable 
at a given initial pressure and temperature with sufficiently high free energy barrier, transition 
can take place in two ways -- (i) increase the temperature until the thermal energy is sufficient to 
overcome the free energy barrier (this is called temperature induced transition and discussed earlier) 
and (ii) increase the pressure until the given initial thermal energy is sufficient to overcome the barrier. 
Increase of pressure favors high density phase and thus decreases the free energy barrier for transition. 
In this section we shall compare these two scenarios and discuss how the pathway of pressure induced transition 
at low temperature differ from temperature induced transition. \\  
\begin{figure*}[ht!]
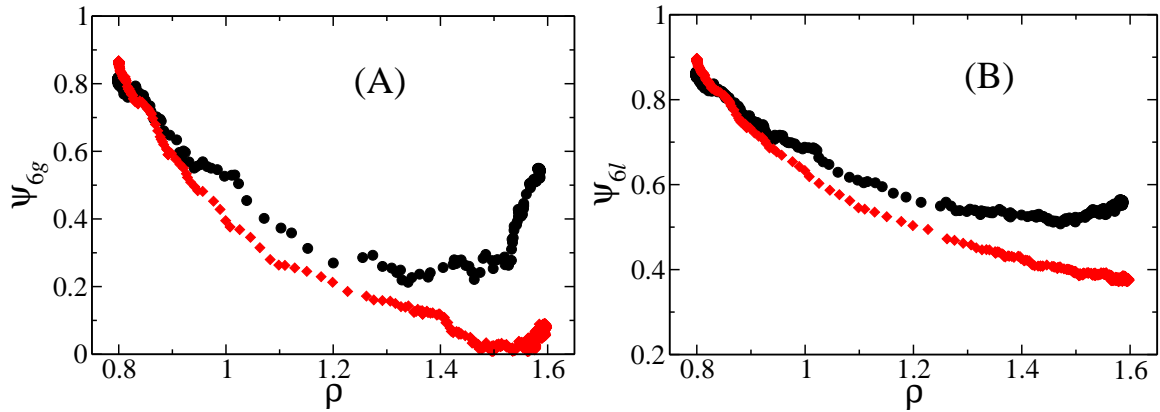

        \subfigure{%
            \label{fig7a}
            \includegraphics[width=0.42\textwidth]{FIG7A.eps}
        }%
        \subfigure{%
           \label{fig7b}
           \includegraphics[width=0.42\textwidth]{FIG7B.eps}
        }\\ 
    \caption{Pathways for pressure and temperature induced transitions in two dimensional 
order parameter (density and global six-fold bond orientational order) space are shown. (A) density ($\rho$) 
and global six-fold bond orientational order parameter ($\psi_{6g}$) space. (B) density ($\rho$) and 
local six-fold bond orientational order parameter ($\psi_{6l}$) space. Black filled circles represent 
the pathway for temperature induced transition and red filled diamonds represent the pathway for pressure 
induced transition.}
   \label{fig7}
\end{figure*}
In Fig. 7 we have compared the pathways for pressure and temperature induced transitions 
in two dimensional order parameter (density and six-fold bond orientational order) space. For temperature induced 
transition the transition temperature is $T$ = 0.105 at $P$ = 0.10. In the case of pressure induced transition 
we have increased the pressure of the system (details 
are mentioned in Section II) at constant temperature $T$ = 0.05 and the observed transition 
pressure is $P\sim0.45$. On increasing pressure transition temperature shifts towards lower temperature. 
In the case of pressure induced transition, we find that different trajectories collapse to different 
metastable disordered states (with slightly different density and order) and transition pressure also 
vary in orders of 0.01. As shown in the figure, contrary to the temperature induced transition where 
the honeycomb solid collapses to high density crystalline oblique solid phase, in the case of pressure 
induced transition the honeycomb solid collapses to a long lived disordered metastable state having very 
small global order parameter value (see Fig. 7A). Disordered metastable phase evolves towards the 
stable state with an extremely slow rate. In spite of the absence of the global six-fold bond orientational 
order, significantly large value of the local six-fold bond orientational order (see Fig. 7B) 
suggests the heterogeneous structure of the metastable phase where significant local six-fold symmetry 
is present but there is no global translation of the local order. \\
\begin{figure}
\includegraphics[width=0.45\textwidth]{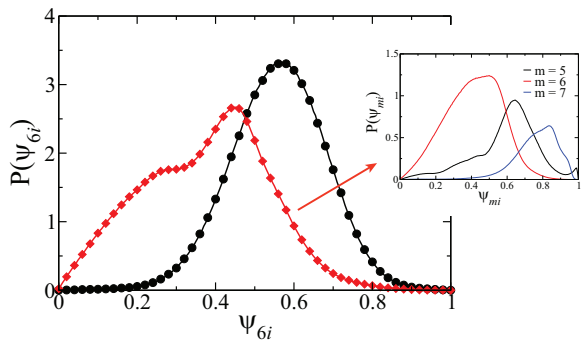}
\caption{Distributions of local six-fold bond orientational order ($\psi_{6i}$) 
for the collapsed structure in the case of temperature induced transition 
(black line distribution with filled circles) and pressure induced transition 
(red line distribution with filled diamonds) are shown. In case of temperature induced transition 
the collapsed structure is oblique solid and in the case of pressure induced transition the 
collapsed structure is disordered solid. In inset, distributions of \textit{m}-fold local 
bond orientational order ($\psi_{mi}$) of particles having coordination number \textit{m} for the pressure 
induced collapsed phase are shown. In the figure the distribution with black line indicates 
the $\psi_{5i}$ distribution of particles 
having coordination number five, red line indicates $\psi_{6i}$ distribution of particles having 
coordination number six and blue line indicates $\psi_{7i}$ distribution of particles having 
coordination number seven. Area under $\psi_{mi}$ distribution gives the fraction of particles 
having coordination number \textit{m}.}
\end{figure}
In order to gain more insight in the microscopic structure of the disordered solid phase, 
in Fig. 8, we have shown the distributions of local six-fold bond orientational order 
($\psi_{6i}$) for both the oblique and the disordered solid phase. 
The emergence of the shoulder in the distribution corresponding to the disordered solid phase 
indicates the presence of heterogeneous local environments. To find the structures of different local 
environments responsible for the broadening of $\psi_{6i}$ distribution towards the lower value, 
in inset we have shown the distributions of $\psi_{mi}$ $(P(\psi_{mi}))$ for 
the particles having coordination number \textit{m}. That is we have computed the distribution of $\psi_{6i}$ for 
only those particles which have coordination number six and so on. The distribution 
is normalized with the total number of particles so that area under curve of $P(\psi_{mi}$) will 
give the fraction of particles having coordination number \textit{m}. We note the presence of significant 
five-cordinated ($\sim30\%$) and seven-coordinated ($\sim18\%$) along with six-coordinated ($\sim50\%$) particles with respective 
symmetries. On heating the metastable disordered solid phase undergoes a transition to the 
stable oblique phase having unimodal \textbf{$\psi_{6i}$} distribution. On increasing pressure the peak 
of the distribution of $\psi_{6i}$ in the oblique solid phase shifts towards lower $\psi_{6i}$ value. \\  
Recently, Kadau \textit{et al}.\cite{5} used molecular dynamics simulations to investigate the 
shock-induced phase transformation of solid iron. Above a critical shock strength the nucleation 
and growth become collective as many small grains start growing simultaneously. However, 
we have observed a single cluster growth scenario in both pressure and temperature induced transitions. 
Collective nucleation and growth scenario indicates the absence of free energy barrier for nucleation. 
Note that shock waves increase both the pressure as well as temperature and increment in both decreases 
the free energy barrier for transition. Above a critical shock strength the free energy barrier might be negligibly small 
and hence many clusters start growing simultaneously. This transition closely resembles with the gas-liquid 
transition at large metastabilty (above kinetic spinodal) where one observes similar collective nature of 
nucleation and growth\cite{26,27,28}.  
In the case of pressure induced transition (or low temperature transition as on increasing pressure 
transition point shifts towards the lower temperature) trapping of the system in a long lived local 
minima indicates that the free energy landscape is rugged. Although one might observe the rugged energy 
landscape in atomic systems also, inclusion of specificity in interaction potential enhances the ruggedness 
of the landscape. The origin of enhanced ruggedness lies in the competition between specific and non-specific 
interactions where local regions might be energetically stabilized by a particular type of interaction but 
system as a whole might not be in global minimum. These local traps in the landscape decreases significantly 
the dynamics of the transition. On increasing temperature the disordered metastable solid phase undergoes a 
transition to the crystalline oblique phase as on increasing temperature system attains sufficient thermal 
energy to overcome local energy barriers and finally undergoes a transition to global free energy minimum.
\section{Conclusion}
In this work we have employed computer simulation techniques to study the solid-solid collapse transition in a 
model molecular system interacting via both isotropic as well as anisotropic interactions. 
In the case of temperature induced transition the metastable honeycomb solid collapses to high density crystalline 
oblique phase. High density oblique solid has two characteristic length scales corresponding to isotropic and 
anisotropic parts of interaction potential. Contrary to the common believe and classical nucleation theory 
that the nuclei of new phase are circular (in 2D) and has same order and density as of bulk stable phase, 
interestingly, we find linear strip-like nucleating clusters having significantly different order and average 
coordination number than the bulk stable oblique phase. Both the order and average coordination number of 
the growing cluster evolve with the size and finally converge to the order of the bulk stable phase. 
In the early stage of growth, the cluster grows as linear strip followed by branched and ring-like strips. 
Complex geometry of growing cluster is a consequence of the delicate balance between two types of interactions 
which enables the dominance of stabilizing energy over the destabilizing surface energy of the cluster. 
The nucleus of the stable oblique phase is wetted by a layer of intermediate order particles which minimizes the surface 
free energy cost for formation of the oblique phase. In the case of pressure induced transition the collapsed state 
is a disordered metastable solid. Pathways of pressure and temperature induced transitions are significantly different. 
These observations shed new light on the complex nature of nucleation and growth in molecular solid-solid transitions as well 
as on the validity of classical nucleation theories in complex molecular systems.

\begin{acknowledgments}
We thank Prof. Surajit Sengupta and Dr. Mantu Santra for helpful discussions. 
We thank DST and BRNS, India for partial financial support of this work. B.B. thanks DST for a J.C. Bose fellowship.
\end{acknowledgments}

\nocite{*}

\bibliography{condmat}

\end{document}